\documentclass{pasj00}
\draft

\SetRunningHead{S. Komugi et al.}{Cold Dust in M33}

\title{Temperature Variations of the Cold Dust in the Triangulum Galaxy M\ 33}
\author{S. Komugi$^{1,2,3}$, T. Tosaki$^{4}$, K. Kohno$^{5}$, T. Tsukagoshi$^{5}$, 
K. Nakanishi$^{1}$, T. Sawada$^{1,2}$, R. Kawabe$^{6}$, H. Ezawa$^{1}$, N. Kuno$^{6}$, S. Onodera$^{6}$,
Y. Tamura$^{6}$, G. W. Wilson$^{7}$, M. S. Yun$^{7}$, K. Scott$^{8}$, T. A. Perera$^{9}$,
J. E. Austermann$^{10}$, D. H. Hughes$^{11}$, I. Aretxaga$^{11}$, K. Tanaka$^{12}$, 
K. Muraoka$^{13}$, R. Miura$^{14}$, and F. Egusa$^{3}$}
\affil{$^1$ALMA Office, National Astronomical Observatory of Japan, 2-21-1 Mitaka, Tokyo, Japan}
\affil{$^2$Joint ALMA Observatory, Alonso de Cordova 3107, Vitacura, Santiago, Chile}
\affil{$^3$Institute of Space and Astronautical Science, Japan Space Exploration Agency, 3-1-1 Yoshinodai, Chuo-ku, Sagamihara, Kanagawa, Japan}
\affil{$^4$Joestu University of Education, 1, Joetsu, Niigata, Japan}
\affil{$^5$Institute of Astronomy, University of Tokyo, 2-21-1 Mitaka, Tokyo, Japan}
\affil{$^6$Nobeyama Radio Observatory, National Astronomical Observatory of Japan, 462-2 Minamimaki, Nagano, Japan}
\affil{$^7$Department of Astronomy, University of Massachusetts, Amherst, Massachusetts 01003, USA}
\affil{$^8$Department of Physics and Astronomy, University of Pennsylvania, Philadelphia, Pennsylvania 19104, USA}
\affil{$^9$Department of Physics, Illinois Wesleyan University, Bloomington, Illinois 61702-2900, USA}
\affil{$^{10}$Center for Astrophysics and Space Astronomy, University of Colorado, Boulder, Colorado 80309, USA}
\affil{$^{11}$Instituto Nacional de Astrofisica, Optica y Electronica (INAOE), Aptdo. Postal 51 7 216, 72000 Puebla, Mexico}
\affil{$^{12}$Department of Physics, Keio University, 3-14-1 Yokohama, Kanagawa, Japan}
\affil{$^{13}$Department of Physical Science, Osaka Prefecture University, 1-1 Sakai, Osaka, Japan}
\affil{$^{14}$Department of Astronomy, University of Tokyo, Hongo, Bunkyo-ku, Tokyo 133-0033, Japan}

\email{skomugi@alma.cl}

\KeyWords{galaxies:ISM}

\Received{$\langle$April 13, 2011$\rangle$}
\Accepted{$\langle$June 5, 2011$\rangle$}
\Published{$\langle$in print$\rangle$}
\begin{document}
\maketitle

\begin{abstract}
We present wide-field 1.1\ mm continuum imaging of the nearby spiral galaxy M\ 33,
 conducted with the AzTEC bolometer camera on ASTE.  We show that the 1.1\ mm flux
traces the distribution of dust with T$\sim 20$ K.  Combined with far-infrared
 imaging at 160 $\mu \mathrm{m}$, we derive the dust temperature
 distribution out to a galactic radius of $\sim 7$ kpc with a spatial resolution of
 $\sim 100$ parsecs.  Although the 1.1\ mm flux is observed
 predominantly near star forming regions, we find a smooth radial
 temperature gradient declining from $\sim 20$ K to $\sim 13$ K
 consistent with recent results from the Herschel satellite.
 Further comparison of individual regions show a strong correlation
 between the cold dust temperature and the $\mathrm{K_S}$ band brightness, but not with the
 ionizing flux.  The observed results imply
 that the dominant heating source of cold dust at few hundred parsec scales 
are due to the non-OB stars, even when associated with star forming regions.
\end{abstract}

\section{Introduction}
Galaxies are known to harbour dust components whose representative
temperatures are warm ($\sim 50$ K) and cold ($\sim 20$ K).  Cold dust
typically represents more than $\sim 90$\% of the total dust mass in
galaxies \citep{devereux90}.  Despite the important role it plays by
providing the dominant formation sites for molecular hydrogen -the fuel
for star formation- relatively little is known about the spatial and
temperature distribution of cold dust either in the Milky Way or in
nearby galaxies.

Far-infrared (IR) flux at $\sim 25-70\ \mu \mathrm{m}$ of nearby galaxies are 
localized to star forming regions, and have temperatures around $\sim
50$ K; this warm component has been attributed to dust heating by UV
photons from massive OB stars \citep{devereux97, hinz04, tabatabaei07a}.  At
longer wavelengths, however, the sources responsible for the dust
heating becomes less evident.  Dust grains with relatively large sizes
$\sim 1\ \mu \mathrm{m}$ can efficiently absorb photons longer than the
UV \citep{draine84}, and may be heated also by stars which are not OB stars
\citep{xu96, bianchi00}.  We will refer to these stars as non-massive stars, 
which may heat large dust grains but are not OB type stars.
 Longwards of $\sim 100\ \mu \mathrm{m}$, where
the flux is sensitive to cold dust around $\sim 20$ K,
the IR flux is known to be composed both from a diffuse disk
component and flux concentrated near star forming regions
\citep{hippelein03} with both components contributing significantly to the total IR flux
\citep{tabatabaei07a}.  Gaining information on the heating sources of such
cold dust, requires that we obtain a temperature distribution map of
various regions in a galaxy.

Dust grains in reality have a distribution of temperatures, but far-infrared and sub-millimeter
observations have suggested that as a whole, they can be characterized
by two (warm and cold) representative temperatures.
Measuring the temperature of dust in a galaxy requires observations at
at least two wavelengths, preferably one on each side of the peak in the
blackbody spectrum, which occurs around 150um for 20K dust.
Far-infrared data are sensitive to warm dust emission below 
$\sim 100 \mu \mathrm{m}$ which further hinders measurements of the
temperature of the cold component \citep{bendo10}.  A key in
circumventing this problem is observations at sub-millimeter
wavelengths, up to $1300 \mu \mathrm{m}$.  Cold dust temperatures have
been derived for numerous galaxies using sub-millimeter data (e.g.,
\cite{chini95}), showing that indeed cold dust below 20 K is ubiquitous
in nearby galaxies, and is distributed on a large scale in star forming
disks.  A major setback of current ground-based sub-millimeter
observations is the lack of sensitivity to faint and diffuse emission,
making large-scale mapping projects of Local Group galaxies difficult.
Existing sub-millimeter observations of galaxies is therefore dominated
by bright galaxies at moderate distances observed at resolution larger
than several hundred parsecs \citep{chini93, chini95, davies99, alton02,
galliano03} which do not resolve individual star forming
regions that are typically $\sim 100$ parsecs in size, or bright
edge-on galaxies \citep{dumke04, weiss08} which do not enable the clear
derivation of temperature distribution within the galactic disk.
 Recent observations by the Herschel satellite have revealed the
effectiveness of deriving temperatures using sub-mm fluxes
\citep{bendo10, kramer10}.

In this paper, we explore the temperature distribution of cold dust in a nearby,
nearly face-on galaxy M\ 33 by means of a large-scale imaging
observation at 1.1\ mm.
The outline of this paper is as follows; section 2 describes the
observation and data reduction strategy, section 3 presents the
obtained image.  Sections 4 presents the temperature distribution in M\
33.  Section 5 further compares the
dust temperature with data at various other wavelengths.  Section 6
summarizes the results and conclusions.

\section{Observation}

Observations of M\ 33 were conducted using the AzTEC instrument \citep{wilson08}, installed on the Atacama Submillimeter Telescope Experiment (ASTE)
\citep{ezawa04, kohno05, ezawa08}, a 10m single dish
located in the Atacama desert of altitude 4800m in Pampa La Bola,
Chile.  Observations were remotely made from an ASTE operation room of the National Astronomical Observatory of
Japan (NAOJ) at Mitaka, Japan, using a network observation system N-COSMOS3 developed
by NAOJ \citep{kam05}.  Basic parameters of M\ 33 are listed in table \ref{m33param}.

\begin{figure}
\includegraphics[width=15cm,keepaspectratio]{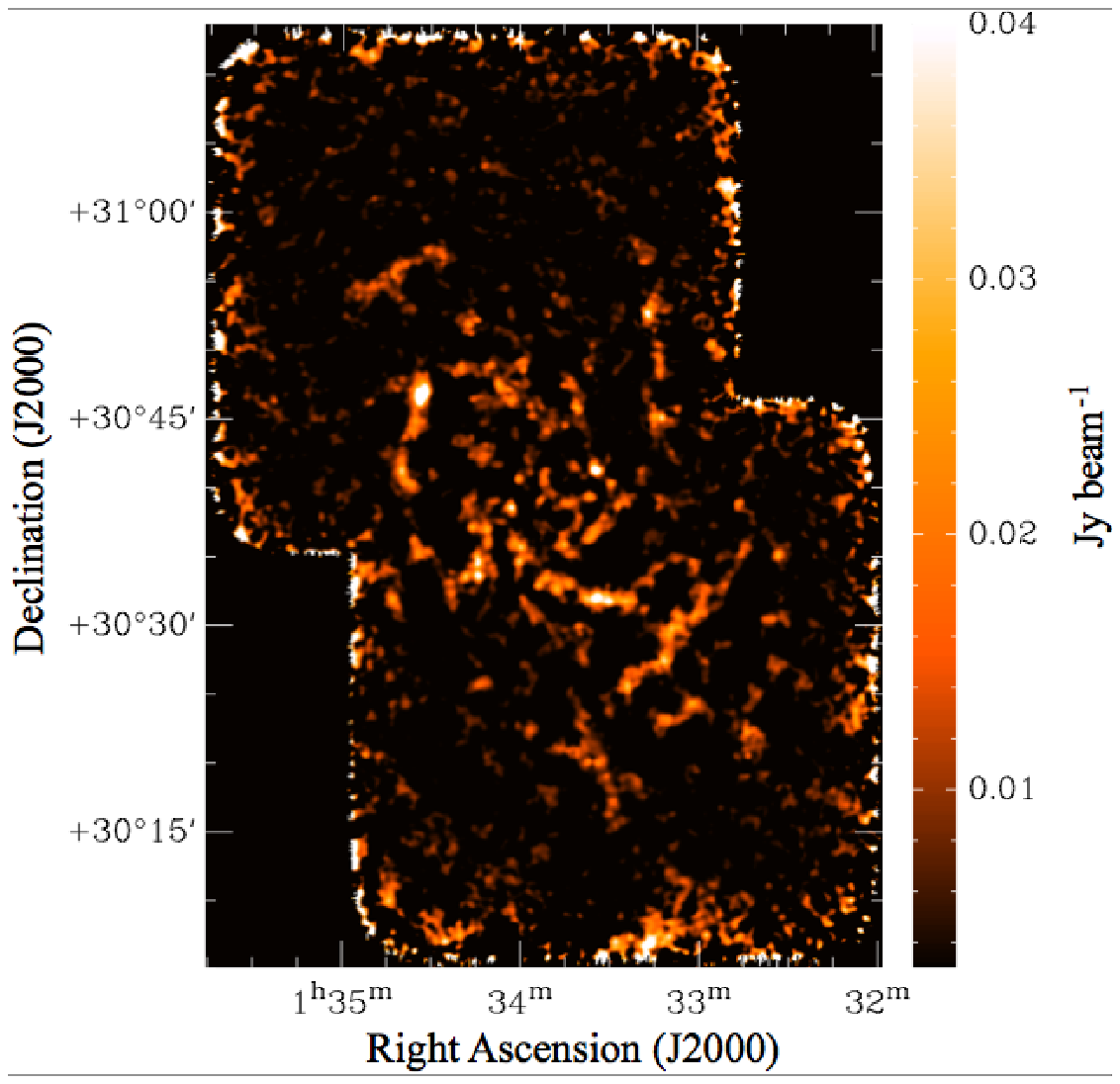}
\caption{M\ 33 observed with AzTEC/ASTE. \label{aztec}}
\end{figure}

Two square fields covering the northern and southern half of M\ 33,
each $30^{\prime} \times 30^{\prime}$ and overlapping by $20^{\prime} \times 5^{\prime}$, where observed at 1.1\ mm (270\ GHz) (see figure \ref{aztec}).
At this wavelength, the array field of view of AzTEC is $7.5^{\prime}$ with a detector FWHM of $28^{\prime \prime}$.  
The field was raster-scanned in the right ascension and declination
 direction alternately, in steps of
 $30^{\prime \prime}$ at a scan speed of $180^{\prime \prime}\mathrm{sec}^{-1}$.
Since the AzTEC detectors are arranged in a hexagonal pattern, this scan step
results in a Nyquist-sampled sky with uniform coverage.

Observations were conducted in July 2007 and August 2008, 
during a total observing time of 40 hours including calibration and pointing
overheads.  Approximately 30 hours were spent on-source.
The zenith opacity at 220 GHz ranged from 0.01 to 0.2, with an average
of 0.06.

Telescope pointing was monitored every two hours using the bright quasar
J0238+166.  The pointing was accurate to $2^{\prime \prime}$ during the observing runs.
Uranus or Neptune was observed twice per night in order to
measure the relative detector pointing offsets and their point spread
functions.
These measurements are also used for determining the absolute flux
calibration.
Relative flux calibration error
is estimated from the flux variation of J0238+166, which was 10\%.  Since
J0238+166 is a variable source, this represents a conservative
overestimate of the calibration error.  Adding in quadrature with the
5\% uncertainty in the brightness temperature of Uranus 
\citep{griffin93}, the total absolute flux calibration error is better than 11\%.
The effective beamsize after beam-smoothing the reduced maps, was $40^{\prime \prime}$.

\subsection{Data Reduction}

Since the raw signals at 1.1\ mm are dominated by atmospheric emission,
we removed the atmospheric signals using an adaptive principal component
analysis (PCA) technique \citep{scott08}.  The PCA method identifies
and subtracts the common-mode signal seen across the bolometer array for each scan.
The PCA calculation is done in the time-domain data, then each scans are mapped and 
co-added to obtain the final image.
Noise is estimated by ``jack-knifing'' the 
individual datasets (that is, multiplying each individual time-stream scans randomly by 
+1 or -1) before the map making.  This cancels out true astronomical signals while
preserving the noise properties of the map, resulting in a realistic representation of the underlying noise.
Details of the PCA method are explained in \citet{scott08}.

The PCA method is optimized for the detection of point sources, by subtracting any extended signals
as atmospheric emission.  Therefore, extended emission from M\ 33 will also be largely subtracted in this 
process.  We use an iterative flux recovery approach called FRUIT to retrieve
extended emission from M\ 33.  The algorithm is briefly explained below.

The idea of the iterative flux recovery approach has already been implemented 
\citep{enoch06} on Bolocam, a bolometer identical to AzTEC, on the Caltech Submillimeter Observatory (CSO).
The FRUIT code used in this study was developed by
the AzTEC instrument team at University of Massachusetts (UMass), and explained in \cite{liu10}.

Structures with scales larger than a several arcmin are removed by the PCA technique because they 
introduce a correlated signal to the detectors which is indistinguishable from atmospheric variations. 
 The residual image created by subtracting the PCA cleaned data from the original dataset therefore
 still contains true astronomical signals which are more extended than what is obtained from PCA cleaning.
The FRUIT algorithm utilizes this residual data, by PCA cleaning them again and
adding the result to the initial result of PCA.  The idea is somewhat analogous to the CLEAN method
\citep{hogbom74} used commonly in interferometric observations at mm
wavelengths.

The observed map of the AzTEC instrument in scan $i$, $M_i$, can be categorized into three
types of signals.  These are the atmospheric signal $A_i$, the true
astronomical signal $S$, and noise $N_i$.  Note that the astronomical
signal is independent of scan.  Although $S$ is the parameter we wish to
derive, we can only construct the best estimator
for $S$ using various algorithms, which we write $\tilde{S}$.  In PCA,
we construct an estimator $\tilde{S_i}$ from $m_i$ (where the lower case
represents the time-domain analog of the upper case maps), and co-add
these $\tilde{S_i}$ to obtain $\tilde{S}$.  The FRUIT algorithm works
iteratively in the following way;

\begin{enumerate}
\item
Construct $\tilde{S_i}$ from $m_i$ using PCA.

\item
Construct $\tilde{S}$ by co-adding $\tilde{S_i}$.  This is where PCA finishes.

\item
Construct a residual, $\tilde{r_i}=m_i-\tilde{s}$.  Ideally, this should
     be astronomical signal free.  In practice, it still contains
     extended emission.

\item
PCA clean the residual $\tilde{r_i}$ and obtain map $\tilde{R}$

\item
Construct the \textit{new} estimator, $\tilde{S'}=\tilde{S}+\tilde{R}$.

\item
Go back to step 3.  The iteration continues until no pixels have signal-to-noise of over 3.5.
\end{enumerate}

\begin{figure}
\includegraphics[width=15cm,keepaspectratio]{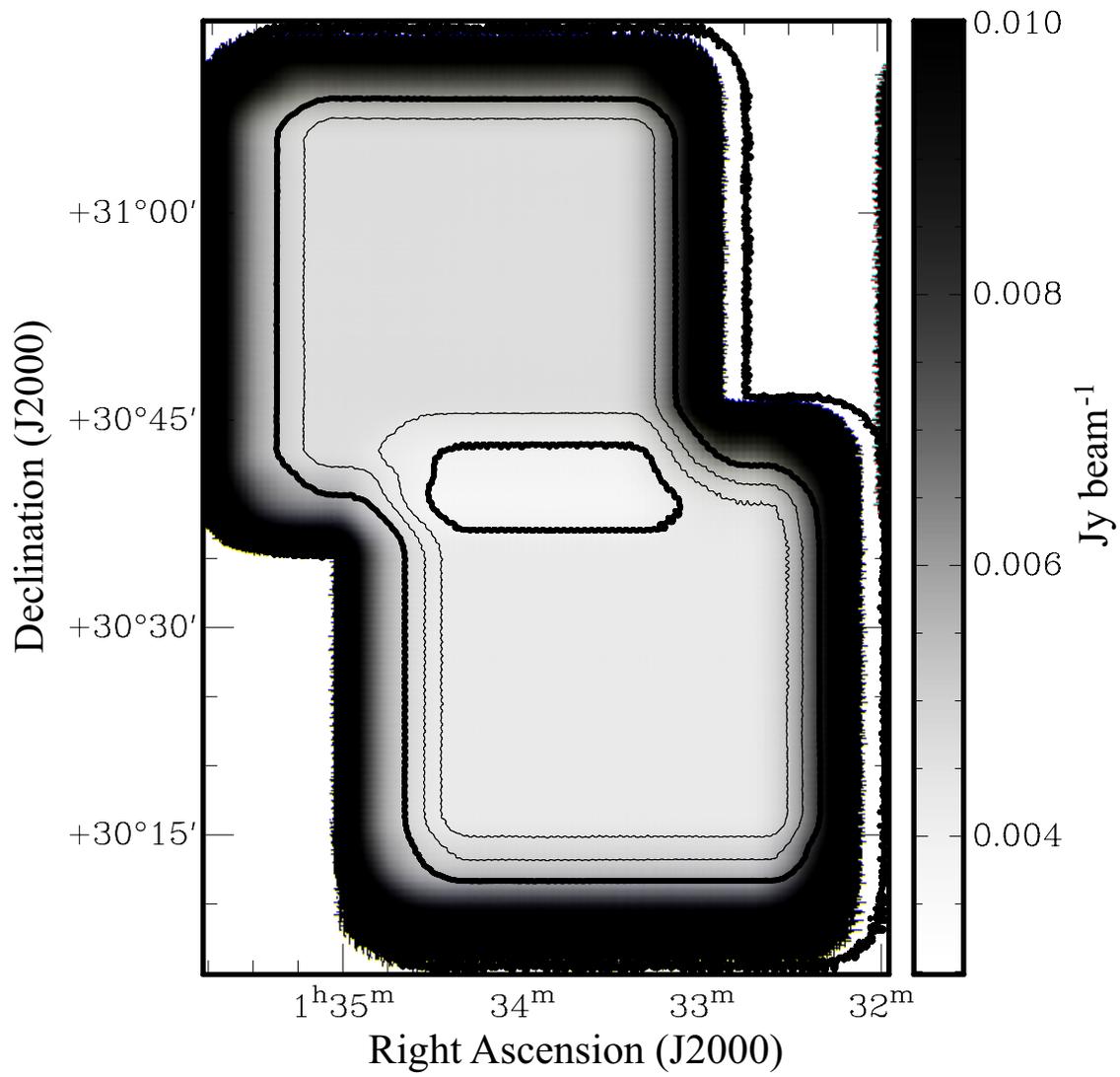}
\caption{Noise map of M\ 33.  The contours are drawn at 
4.0, 4.5, 5.0 and 6 $\mathrm{mJy\ beam^{-1}}$, from the center outwards. \label{noise}}
\end{figure}

The 1.1\ mm image obtained from the FRUIT algorithm is shown in figure \ref{aztec}, and the
noise map obtained by jack-knifing the residual maps in figure \ref{noise}.

This algorithm was tested by embedding two dimensional Gaussian components with FWHM of $1^{\prime}$, 
$2^{\prime}$, and $3^{\prime}$ and various peak
intensities in the original data, and measuring the fraction of the
extended flux recovered with this iterative cleaning method.  We find
that this technique retrieves 90\% of the flux in extended structures up
to $2^{\prime}$, and $\sim 60\%$ for structures with angular sizes of
$3^{\prime}$.  Results of this test are summarized in table \ref{embed}.  The results are
consistent with the same test done in \citet{shimajiri10}, and shows that the 1.1\ mm flux
obtained at the scale of few arcminutes are reasonable.

The northern and southern halves of the galaxy were mapped using FRUIT individually, and then 
merged by weighting the overlapping region by the inverse square of their noise level.
Regions with less than 50\% coverage in each of the northern and southern halves were
clipped from further analysis.

The final maps are constructed at $6^{\prime \prime}$ per pixel.
  The root mean square
(r.m.s.) of the resulting noise map is
 $3.7-4.0 \ \ \mathrm{mJy\ beam^{-1}}$ in the central $\sim 100\ \mathrm{arcmin^{-1}}$, 
increasing to $4.3-4.7 \ \ \mathrm{mJy\ beam^{-1}}$ in the outermost
regions where scientific analysis is conducted (see figure \ref{noise}).

\begin{figure}
\includegraphics[width=6cm]{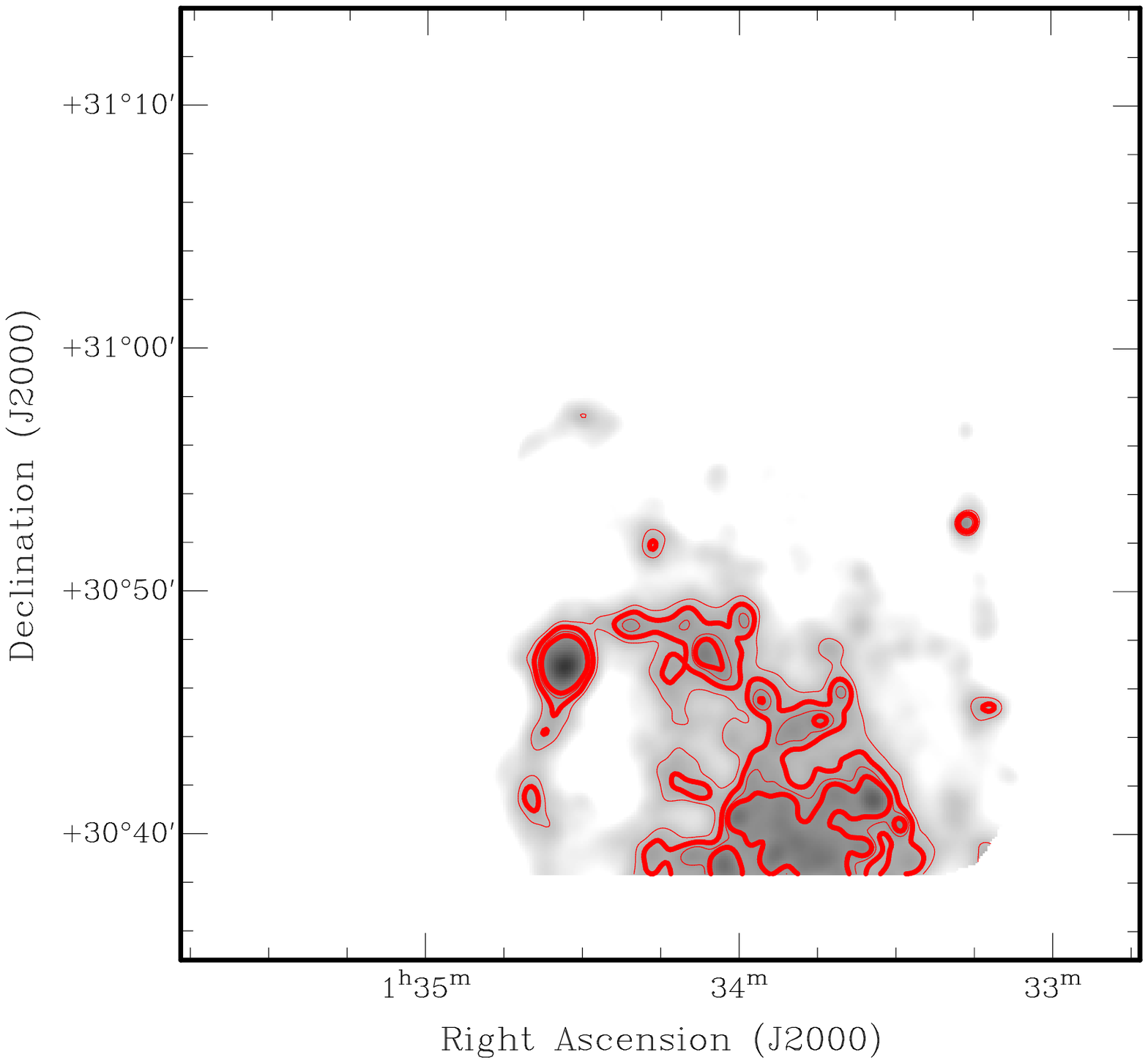}
\includegraphics[width=6cm]{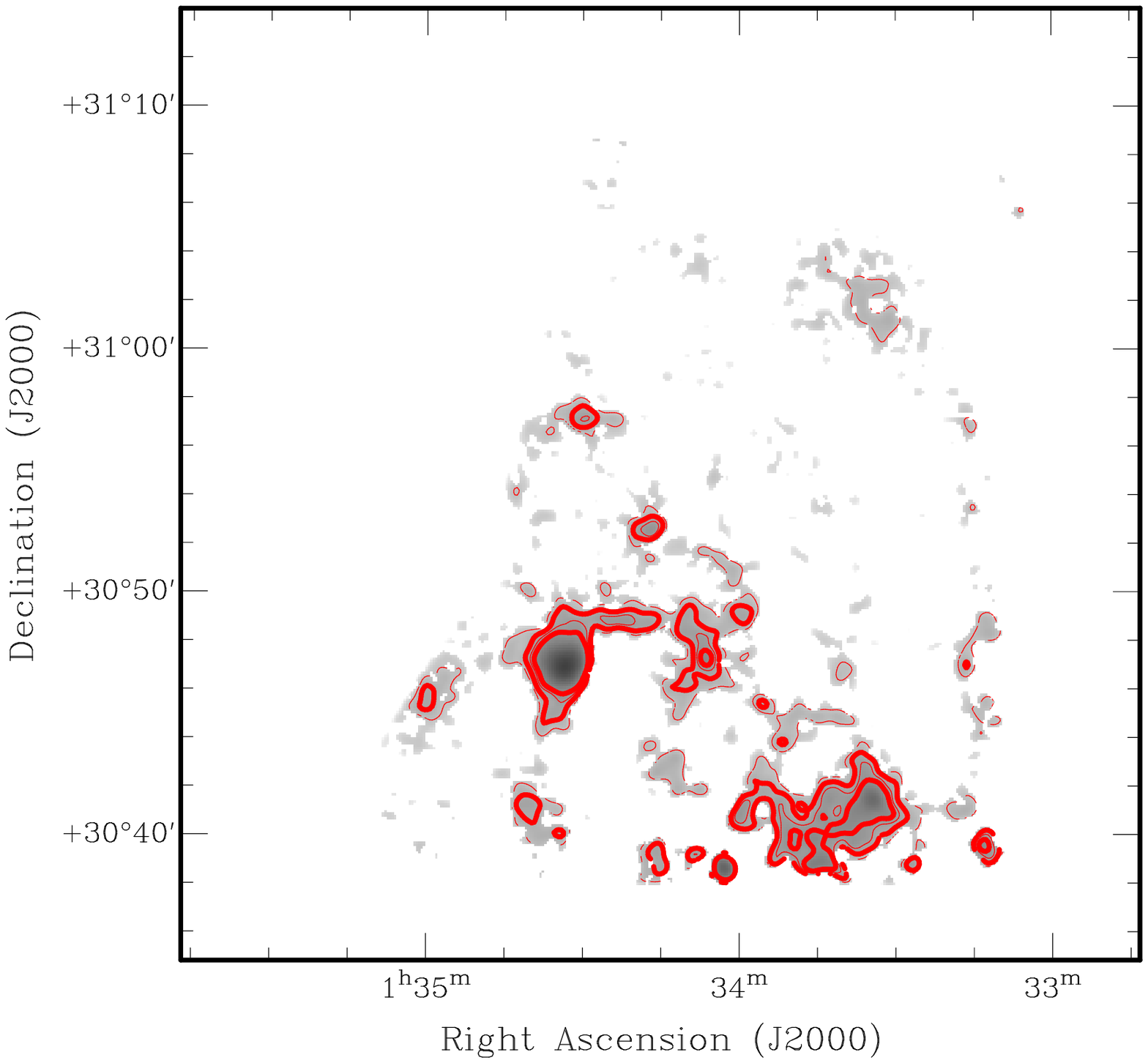}
\caption{Left: Input Spitzer MIPS 160 $\mathrm{\mu m}$ image, scaled to AzTEC fluxes.  Contours are drawn from 15 $\mathrm{mJy\ beam^{-1}}$ in steps of 5 $\mathrm{mJy\ beam^{-1}}$. Right: Output Spitzer MIPS 160 $\mathrm{\mu m}$ map, after applying PCA and FRUIT.  Contour steps are same as the left panel.
The southern most edge with a coverage of $<$ 50\% of the peak is clipped.
\label{sim_image}}
\end{figure}

As a further test of how much extended emission is recovered in the reduction,
we apply the same technique as in \citet{liu10}, by applying the FRUIT algorithm
to Spitzer $160\ \mathrm{\mu m}$ data (see section 4.1) to see the level of flux loss.
The test was done on the northern half of M\ 33.
The Spitzer data were first scaled so that the flux level was similar to AzTEC data, 
to ensure that the same filtering was applied.  The scaled data were then added to the AzTEC 
noise map in the time-stream domain, and then the PCA and FRUIT algorithms were applied.
The left and right panels of figure \ref{sim_image} show the MIPS map prior to and following the PCA+FRUIT reduction,
 respectively
.  The most extended features at scales over $5^{\prime}$ are clearly
subtracted out, but smaller scale features $< 5^{\prime}$ are retrieved.  The total flux recovered by 
FRUIT was 54\% of the input flux, but when restricted to pixels with S/N over 2 in the FRUITed image,
97\% of the total flux was recovered.

\begin{figure}
\includegraphics[width=15cm]{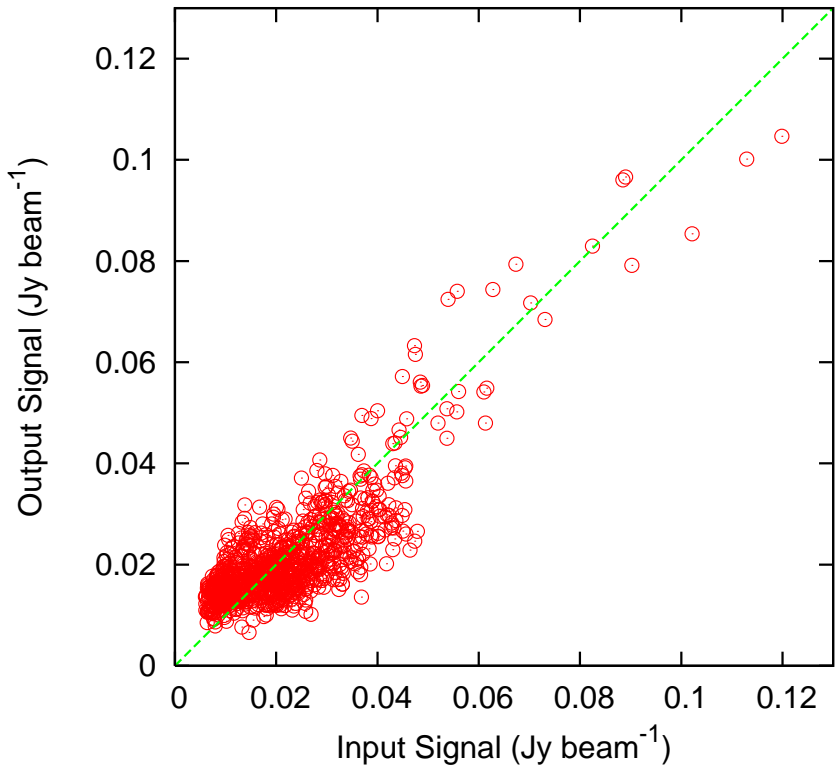}
\caption{Input versus the output flux after FRUIT flux retrieval, for the MIPS 160 $\mathrm{\mu m}$ map. \label{sim_data}}
\end{figure}

Figure \ref{sim_data} shows the output flux as a function of input 
flux, only for pixels with a local S/N over 2 in the output map.  
The 1$\sigma$ standard deviation of the points in figure \ref{sim_image}
around the line of unity is 44\%.  Although this is a simulation using MIPS images, 
we regard this as a representative
uncertainty of the flux measurements in the AzTEC map.  It is likely to be an overestimate of
scatter, because millimeter emission is often more clumpy than at far-IR; clumpy structures are 
retrieved by FRUIT more efficiently than smooth features.
Added in quadrature with the absolute calibration error of 11\% (section 2), the pixel based 
flux uncertainty of the M\ 33 map is 45\% after PCA+FRUIT reduction. 
  The simulation implies that smaller
scale structures (typically smaller than $3^{\prime}$) retrieved in the PCA+FRUIT are accurate to this uncertainty and recovered to 97\% if detected at 
S/N over 2, but larger scale features are subtracted out and are not quantified by the 45\% uncertainty.

\section{Results}
\begin{figure}
\includegraphics[width=10cm]{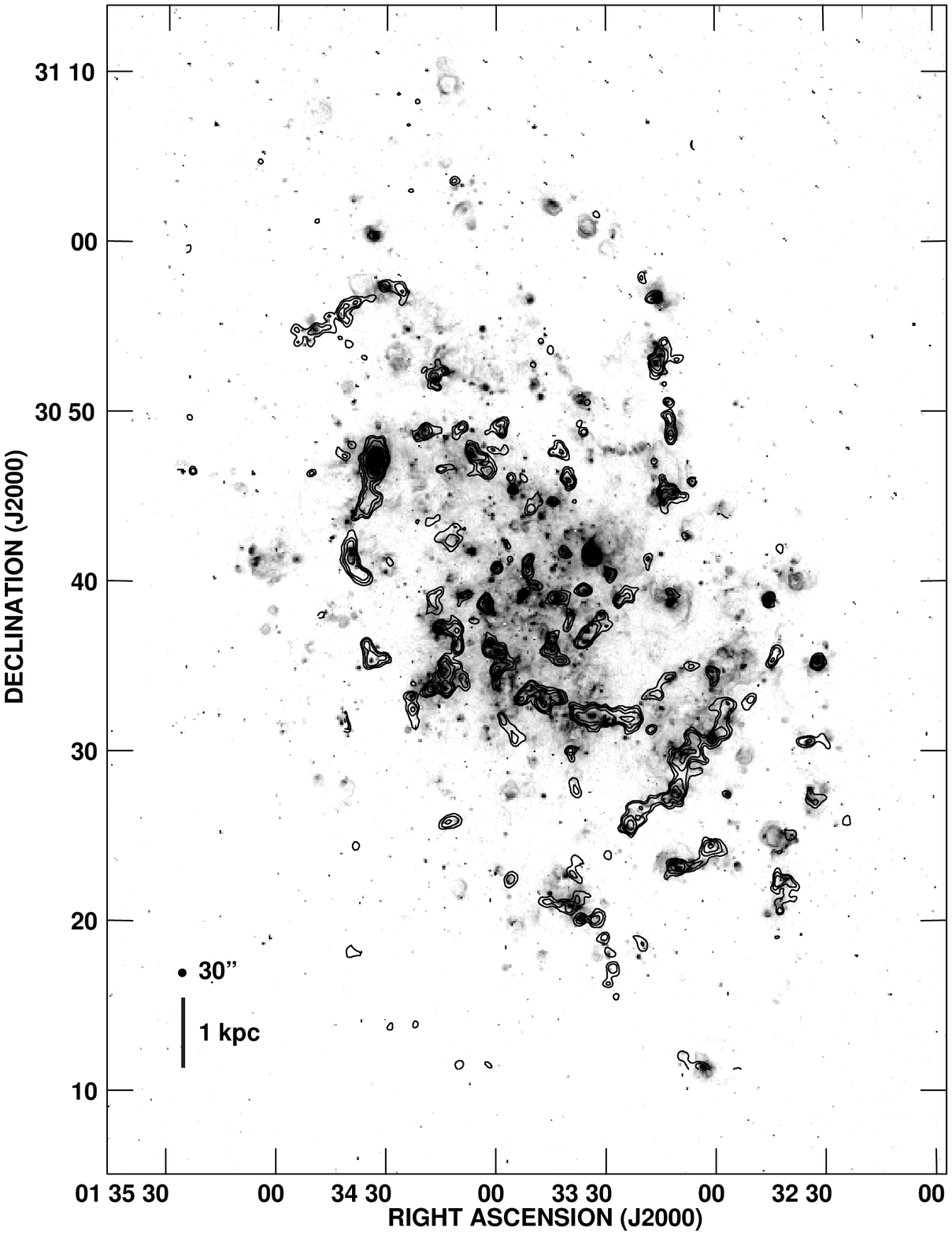}
\caption{1.1\ mm contours overlayed on H$\alpha$ image from \citet{hoopes00}.
Contours are drawn at 9.0, 13, 18, and 25 $\mathrm{mJy\ beam^{-1}}$. \label{contour}}
\end{figure}

Figure \ref{contour} shows the obtained 1.1\ mm contour map of M\
33, overlaid on a narrow-band H$\alpha$ image.  At the distance of 840
kpc for M\ 33 \citep{freedman91}, the angular resolution of
 $40^{\prime \prime}$ corresponds to $160$ pc.

A clear spiral pattern is observed, coincident with the optical arms.  
The 1.1\ mm continuum is detected out to a radius of $\sim 7$ kpc, which is
coincident with the edge of the star forming disk \citep{verley09}.
Almost all of the 1.1\ mm clumps are observed to be associated with star
forming regions as seen in the H$\alpha$.  This is consistent with
previous observations in the far-IR \citep{devereux97, tabatabaei07a}.  An extended disk is not observed,
however, presumably due to the subtraction of extended features during
the data reduction process.
  Although the global 1.1\ mm morphology is
similar to that seen in H$\alpha$, the contamination of free-free
bremsstrahlung to 1.1\ mm is expected to be negligible.  

For optically thin HII regions observed at frequencies higher than few
Hz, the free-free emission can be written in the form;
\begin{equation} \label{freefreeeq}
F_\nu^\mathrm{ff} \propto \mathrm{N_e T_e}\nu^{-0.1}
\end{equation}
where $\mathrm{N_e}$ and $\mathrm{T_e}$ are the electron number density
and electron temperature of the HII region, respectively.
\citet{tabatabaei07b} observed M\ 33 at $3.6$ cm ($8.3$ GHz), where the flux is
dominated by synchrotron radiation and thermal free-free emission.
From \citet{tabatabaei07b}, the typical $3.6$ cm
emission seen in the HII regions have peak fluxes of 
$\sim 10 \ \mathrm{mJy \ beam^{-1}}$.
Taking into account the typical thermal fraction of HII regions as $\sim 50\%$ (\cite{tabatabaei07c}),
and scaling by equation \ref{freefreeeq}, we obtain the $1.1\ $mm flux estimate of free-free
emission at HII regions, of $3.5 \ \mathrm{mJy \ beam^{-1}}$.
This is barely detectable with the r.m.s noise of our observations, which goes down to $3.7\ \mathrm{mJy\ beam^{-1}}$.
For regions which we will discuss in the following sections, the $1.1\ $ mm flux is
typically $\sim 30\ \mathrm{mJy\ beam^{-1}}$.  Therefore, the free-free contribution to the $1.1\ $ mm flux is 
at most around $12 \%$.  We will therefore assume hereafter that the $1.1\ $mm flux is the pure flux from the cold dust
continuum.

\begin{figure}
\includegraphics[width=10cm]{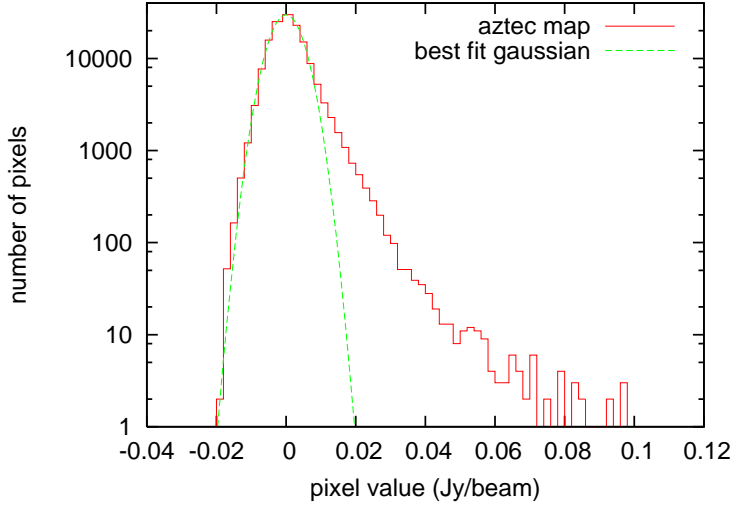}
\caption{The probability density function (PDF) of the AzTEC map.  The red histogram shows the PDF, and the dashed green line
shows the best fit gaussian using pixels below $15\ \mathrm{mJy\ beam^{-1}}$.   \label{pdf}}
\end{figure}

The number probability density function (PDF; distribution of fluxes values in each pixel) of the
whole AzTEC map was used to obtain a realistic estimate of the total 1.1\ mm flux.  This was done 
in place of the normal aperture photometry, in order to circumvent the possibility of including 
significant negative pixels in the total flux which can be added by the PCA method.  It can also deal 
with large scale correlated noise, which may have systematic effects on the total flux.  
Figure \ref{pdf} shows the PDF of the AzTEC map, clipped where the r.m.s. noise level exceeded 
$6\ \mathrm{mJy\ beam^{-1}}$, within which we have done any analysis.
The faint end of the distribution, dominated by noise, is well fitted by a gaussian with a dispersion of $6.0\ \mathrm{mJy\ beam^{-1}}$.
This shows that any negative pixels introduced by the PCA method and FRUIT retrievals, are offset by any positive values introduced, so that
the sum of all the pixels in the AzTEC image still represents the true observed total flux of M33 reasonably well.
The sum of all pixels with values larger than $-15\ \mathrm{mJy\ beam^{-1}}$, or about -3 $\sigma$, was 10.0 Jy.  Including all pixels
below $-15\ \mathrm{mJy\ beam^{-1}}$, the total flux is 9.9 Jy.
No pixels had values lower than $-20\ \mathrm{mJy\ beam^{-1}}$.  
The flux retrieval simulation using MIPS measurements in the previous section showed that FRUIT 
retrieved $\sim$ 50\% of the total flux when no restrictions to the signal-to-noise ratio are given,
so the intrinsic $1.1\ $ mm flux is be estimated to be $\sim$ 20 Jy, which can be regarded as the flux upper limit.
Therefore, a range of $9.9 - 20$ Jy gives a reasonable estimate of the total observed $1.1\ $ mm flux of M\ 33.

\section{Temperature Distribution}

\subsection{Spitzer MIPS 160um Data}

We retrieved the $160 \ \mu \mathrm{m}$ Multiband Imaging Photometer
(MIPS: \cite{rieke04}) datasets (AORs 3648000, 3648256, 3649024, 3649280, 3650048, 15212032 and 15212288)
 of the Basic Calibrated Data (BCD) created by Spitzer Science Center (SSC) pipeline from the Spitzer Space 
Telescope \citep{werner04} data archive.  Individual frames were processed using MOPEX \citep{makovoz05} version 18.4, 
rejecting outliers and background-matching overlapping fields.  Each mosaics of dimension $100^{\prime} \times 20^{\prime}$ were 
co-added using WCS coordinates.  The co-added image was then background-subtracted by selecting several regions away from M\ 33 (by more than
$30^{\prime}$), fitting, interpolating and subtracting a first order plane.  The
1$\sigma$ background noise level was measured at several regions well away from the disk of M\ 33, which
was at most $0.9 \ \mathrm{mJy \ Str^{-1}}$ per $8^{\prime \prime}$ pixel.

The flux error of $160\ \mu \mathrm{m}$ is the sum of the error in zero-point
magnitude of $1.5\%$, and in the conversion from instrumental units to
$\mathrm{MJy \ Str^{-1}}$ of $12\%$ (both from \textit{Spitzer} Science Center
Home-page).  Additionally, the background noise level of 
$0.9 \ \mathrm{MJy \ Str^{-1}}$ must be taken into
account.  The weakest features for which the temperature was derived (in
the outermost disk) has a $160\ \mu \mathrm{m}$ intensity of 
$\sim 10\ \mathrm{MJy \ Str^{-1}}$, so background noise can account for up to
$\sim 10\%$.  All errors taken into account, the pixel-to-pixel $160\ \mu \mathrm{m}$
flux should be accurate to $\le \sqrt{1.5\%^2+12\%^2+10\%^2}=16\%$.
The angular resolution (FWHM) of the point spread function is $40^{\prime \prime}$.

\subsection{Global Temperature}
The total $160 \ \mu \mathrm{m}$ flux measured within a $60^{\prime}$ aperture (regions inside the aperture but outside
the observed region were masked by null) was $\sim 1900\ \mathrm{Jy}$, well
within the uncertainties of the measurements by \citet{hinz04}, $2054\pm411\ \mathrm{Jy}$, so the MIPS data reduction presented in the
previous section is consistent with other studies in the literature.  In the 
SED fit explained below, we use the $160\ \mathrm{\mu m}$ value by \citet{hinz04}.  
Far-IR flux measurements at other wavelengths are listed in table \ref{sedflux}.
The far-IR SED is known to be successfully represented by two-component modified blackbodies shortwards of
500 $\mathrm{\mu m}$ in various regions within the galaxy \citep{kramer10}, and we assume a
two-component spectrum also.  This assumption can break down if there is significant contribution
from the free-free thermal component at the long wavelengths (but see section 3), or significant 
contribution from non thermal-equilibrium dust (very small grains with small heat capacity) exist at the shortest
wavelengths $\sim$ 24 $\mathrm{\mu m}$ \citep{hippelein03}.

\begin{figure}
\includegraphics[width=10cm,keepaspectratio]{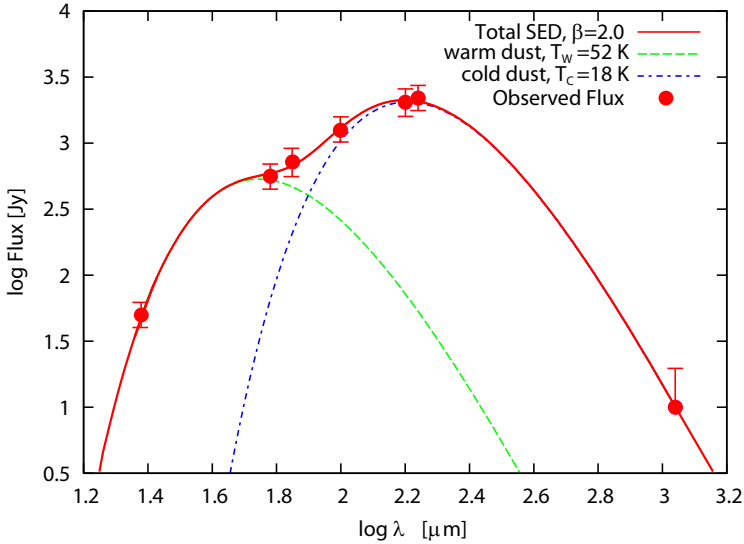}
  \caption{Two component fit to the observed global SED of M\ 33.  The 
green dotted curve is the warm dust component with temperature
 $\mathrm{T_w}=52\pm7$ (K), and the blue dash-dotted curve is the cold dust
 component with temperature $\mathrm{T_c}=18\pm3$ (K), using the observed 1.1\ mm flux of 10 Jy.
The red solid curve is the sum of the two components.  Errors in the infrared fluxes are
typically 20\%, and are taken from \citet{hippelein03} and \citet{hinz04}.  For the 1.1\ mm flux,
explanation of errors are given in section 3. \label{sed}}
\end{figure}

The flux $F_\nu$ at various IR bands were fitted with a two-component modified blackbody
model of the form
\begin{equation}\label{twocomponenteq}
F_\nu = \nu^\beta \{aB_\nu(\mathrm{T_w})+bB_\nu(\mathrm{T_c})\}
\end{equation}
where $a$ and $b$ are constants, $\beta=2.0$ is the assumed dust emissivity index
, and $B_\nu(\mathrm{T})$ is a Planck function at temperature T.
  The chi-square fitted
result is shown in figure \ref{sed}.  We obtained 
$\mathrm{T_w} = 52 \pm 7$ K, $\mathrm{T_c} = 18 \pm 3$ K and $\chi^2 = 0.5$, where $\chi^2$ is the
reduced chi-square value.  For a single component modified blackbody with varying $\beta$, we obtain
$\mathrm{T} = 56 \pm 10$ K, $\beta = 0.2 \pm 0.5$, and $\chi^2 = 13$, so this high
reduced chi-square value rules out a single component 
modified blackbody spectrum.  Considering the uncertainty of $\beta$ and that
many previous studies have derived $\beta \sim 2$ \citep{knacke73, draine84, gordon88,
mathis89, chini93, kruegel94, lis98, dunne00, dunne01, hill06, kramer10}, we keep $\beta=2.0$
constant hereafter.

Our derived global cold dust temperature $\mathrm{T_c}$ should be considered an
slight overestimate, since extended features are subtracted in the data
reduction (see section 2.1).  If we use the corrected 1.1\ mm flux estimate of 20 Jy, we
obtain a cold dust temperature of $16 \pm 2$ K.  This temperature
and the corrected 1.1\ mm flux is consistent with extrapolations of the far-IR
SED to 1.1\ mm in previous measurements \citep{hippelein03} and the recent
value derived from Herschel \citep{kramer10}.

Using the observed 1.1\ mm flux of 10 Jy and cold dust temperature of 18 K, and assuming optically thin
dust and a dust emissivity of $\kappa = 0.114 \ \ \mathrm{m^2 \
kg^{-1}}$ \citep{ossenkopf94}, we obtain a total observed dust
mass of $1 \times 10^6 \ \ \mathrm{M_\odot}$.  For an estimated intrinsic flux of 20 Jy and the
resulting temperature of 16 K, the dust mass becomes $2.8 \times 10^6 \ \ \mathrm{M_{\odot}}$.

The neutral atomic mass is $1 \times 10^9 \ \ \mathrm{M_\odot}$ \citep{newton80} and the 
molecular gas mass is estimated to be $2.6 \times 10^8 \ \ \mathrm{M_\odot}$
from CO observations \citep{heyer04}, so the total dust-to-gas ratio is derived to be $\sim 500 - 1300$.

An important indication from figure \ref{sed} is that the two
different temperature components both contribute significantly at
wavelengths around and below $100\ \mu \mathrm{m}$, whose fluxes combined with fluxes
$\ge 100 \ \mu \mathrm{m}$, are used commonly to discuss the temperatures
of the cold dust component.  This can significantly overestimate
the cold dust temperature.  \citet{verley09} gave a cold dust temperature
declining from $\sim 25\ $K to $\sim 21\ $K in M\ 33 using the color temperature of  
$70\ \mu \mathrm{m}$ combined with $160\ \mu \mathrm{m}$, but at $70\ \mu
\mathrm{m}$, figure \ref{sed} shows that the two dust components
contribute almost equally.  \citet{bendo10} and \citet{kramer10} also show that wavelengths
shortwards of $100 \mu \mathrm{m}$ are contributed significantly by star formation.
Although color temperatures derived using only the two bands $\sim 160\ \mu \mathrm{m}$ and $1100 \ \mu \mathrm{m}$ can
still suffer from variations in $\beta$ (see section 4.5), it gives a more reasonable representation of the colder component
 and its characteristic temperature $\mathrm{T_c}$ compared to attempts using shorter wavelengths.

\subsection{Color Temperature}
The color temperature was derived using 
\begin{equation}\label{eqcolortemp}
\mathrm{R} \equiv \frac{F_{160\ \mu \mathrm{m}}}{F_{1100\ \mu \mathrm{m}}}=\left( \frac{1100}{160} \right)^{-\beta} \frac{B_{160}(T)}{B_{1100}(T)}
\end{equation}
which is obtained from equation (\ref{twocomponenteq}) at two
wavelengths, for a single temperature component.  Equation
(\ref{eqcolortemp}) was solved numerically to obtain the correspondence
between observed flux ratio and the color temperature, and interpolated to the observed ratios.
Uncertainties were estimated from the flux uncertainties in the 1.1\ mm flux of 45\% (section 2) and
160 $\mathrm{\mu m}$ of 16\% (section 4.1), which added in quadrature results in an flux ratio uncertainty estimate of 48\%.
At temperatures of 10 K, 18K, and 20 K, the flux ratio $\frac{F_{160\ \mu \mathrm{m}}}{F_{1100\ \mu \mathrm{m}}}$ is 20, 120, and 160, 
corresponding to color temperature uncertainties of 0.9 K, 2.3 K, and 2.8 K, respectively.


In the actual derivation, the region corresponding to the observed AzTEC
field was cut out from the $160\ \mu \mathrm{m}$ image, and further re-gridded to the 
same pixel size ($6^{\prime \prime}\mathrm{pix}^{-1}$).  The \textit{Spitzer} field of view did not include the
northeastern and southwestern edge regions, and are excluded from temperature analysis.

\begin{figure}
   \includegraphics[width=10cm,keepaspectratio]{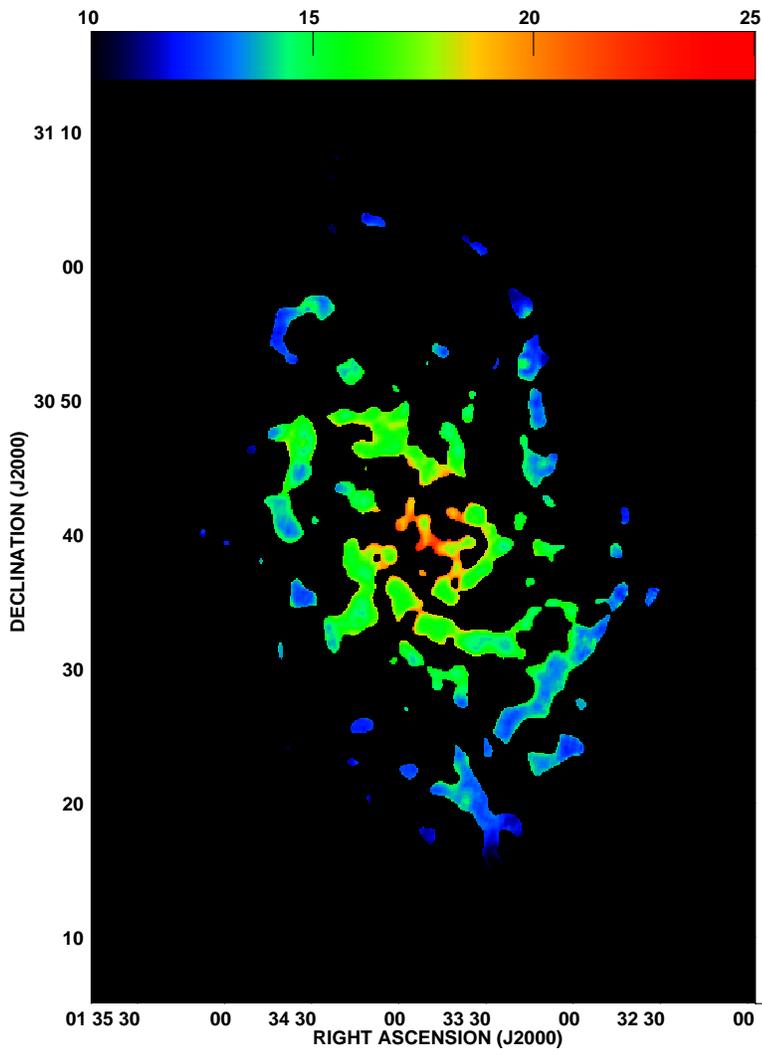}
  \caption{The cold dust temperature map of M\ 33.  Units are in
 $\mathrm{T_c}$(K). \label{temp}}
\end{figure}

The two images were then divided pixel-by-pixel, using only those pixels where $1.1\ $mm
fluxes exceeded $6.0 \ \mathrm{mJy \ beam^{-1}}$ (the characteristic 1$\sigma$ noise level
 of the map, but $\ge 1.5\sigma$ for most cases. See figure 2.) and $160 \ \mu \mathrm{m}$
flux exceeded the $1\sigma$ noise level of $0.9\ \mathrm{mJy \ str^{-1}}$,
to ensure that the flux ratios are not affected by noise (virtually no cuts were made for $160 \ \mu \mathrm{m}$ data using this 
criterion, however; all regions where the map satisfied the $1.1\ $mm criterion had $\sim 10 \sigma$ in the \textit{Spitzer} map).
The flux ratios were then converted into temperature
using the numerical solution to equation \ref{eqcolortemp}. The obtained cold dust 
temperature map is shown in figure \ref{temp}.  The color temperature derived for the global flux was
21 K for the total observed flux of 10 Jy, and 18 K for the estimated total intrinsic flux of 20 Jy.  These values are 
 within the uncertainties presented for temperature derivation using two-component modified blackbody fits in section 4.2.

A prominent feature evident from the temperature map, is that the local
variation of temperature at small scales of several arcminutes
($1^{\prime}=240$pc) is small compared to the global variation seen at
kpc scales, namely the temperature decrease observed from the center
outwards.  Only a few intense star forming regions are found to have
temperatures slightly higher than the surrounding dust, and the bulk of
dust seems to be at temperatures that follow a global trend.
  This is unexpected in the case where recent star formation
governs the dust temperature, because star formation varies its
intensity at small scales.  The smooth distribution of temperatures
despite the range of star formation intensities in these regions, show
that the observed concentration of 1.1\ mm flux nearby star forming regions is
simply the result of more dust being there, not a higher temperature.

For each of the arcminute-sized clumps in the map, the temperature seem
to be higher at the edges compared to the inner regions.  This may be
caused by the data reduction procedure, since the flux retrieval
fraction of extended features decrease with increasing source size.  At
large scales, the 1.1\ mm flux is underestimated, giving a higher
temperature.  

Figure \ref{gradient} shows the radial temperature gradient, constructed
from figure \ref{temp} averaged over radial bins of $90^{\prime
\prime}$ width, which is chosen to mitigate the effect of
temperature variations at the edges of arcminute structures.  The
declining temperature gradient is clearly apparent.  Only a few cold dust
temperature gradients have previously been found (e.g., \cite{kramer10}).  It is
important to note that this gradient is found even though the 1.1\ mm
emission is found predominantly where star formation is active; even for
dust associated with star forming regions at $\sim 100$ pc scales, M\ 33 has a smooth
temperature gradient which varies globally.

\subsection{Correction for Extended Emission}\label{gradient}
As discussed in section 2.1, the total flux retrieved by FRUIT for \textit{Spitzer}
$160 \mu \mathrm{m}$ data was 54\%.  We can also assumed this to be
the fraction of flux lost in the 1.1\ mm data analysis, i.e., $\sim$ 10 Jy (section 3).
Such 1.1\ mm flux would reside in extended structures typically over $5^{\prime}$, 
corresponding to $\sim$ 1.2 kpc.   
Extended dust disks of kilo-parsec size have been detected in the
sub-millimeter range \cite{meijerink05}, and the existence of such a
structure in M\ 33 would be the obvious candidate of structures which
may be lost in the data processing.  If this extended dust disk had a
radially declining flux distribution, more 1.1\ mm flux would be lost 
in the central regions, giving a higher 
dust temperature when compared with $160 \mu \mathrm{m}$ data.  This can
mimic a temperature gradient.  Below we attempt to estimate that this effect
can have on our derived temperature gradient.

We assume here that the total flux missed by 1.1\ mm data analysis is
10 Jy, and that it forms a smooth exponential disk as in \citet{meijerink05},
with a scale radius of 2.4 kpc as derived from both
the $160 \mu \mathrm{m}$ \citep{verley09} and CO observations
\citep{heyer04}.  An exponential disk gives more extended flux to the central region,
thus can be considered to be the worst case scenario, while giving a physically plausible model.
 The dust temperature was recalculated by accounting for
this model in the 1.1\ mm flux for the flux ratio R in equation
(3).  The resulting temperature profile is shown as green
dashed lines in figure \ref{gradient}.  Although the temperature
decreases in the central few kpc by 1-2 K, a radial decline is still apparent.

\subsection{Variation in $\beta$}
Properties of dust grains (i.e., size distribution and optical
properties) are usually expressed in terms of their dust grain
emissivity index $\beta$.  Although we have used a fixed $\beta$ for our
analysis, grain properties and their size distribution may vary within a galaxy thus changing $\beta$;
this corresponds to the case where a two temperature decomposition of the 
SED is not valid, and must include a range of temperatures.  This is expected theoretically
 \citep{li01, dale02}, and a wide range of dust
 temperatures have actually been observed in our Galaxy (\cite{reach95}).

Variations in $\beta$ can alter the derived cold dust temperature.  
In particular, a radial variation in $\beta$ may be able
to mimic a temperature gradient.  For the temperature gradient in figure \ref{gradient} to
be entirely explained instead by a constant temperature at 20 K and variation in $\beta$,
$\beta$ must be $\sim 2$ in the central regions and decreasing to $\beta \sim 1$
 in the outer several kilo-parsecs.  Although variations in $\beta$ are known to be $2 \pm 0.6$
\citep{hill06} in the Milky Way, systematic variations with respect to
regions within a galaxy have not been reported.  Furthermore, low values
of $\beta$ around 1 - 1.5 are typically found near individual proto-stars
\citep{weintraub89, knapp93, williams04}.  It is unlikely that the
characteristics of dust found in such circumstellar environments dominate
over the outer region of M33 at kilo-parsec scales.  Although small radial variations 
in $\beta$ cannot be ruled out with our current dataset, 
a global temperature gradient seems to more plausibly explain the
flux ratio gradient between 1.1 mm and 160$\mu \mathrm{m}$.

\section{Temperature - $\mathrm{K_S}$ - H$\alpha$+24$\mu \mathrm{m}$ Correlation}
In order to assess the heating sources of cold dust within sub-regions of
the galaxy, we performed aperture photometry around individual HII
regions, and compared the dust
temperature with properties at various wavelengths.

The $\mathrm{K_S}$ band ($2.1 \mu \mathrm{m}$) can be used to trace the 
stellar distribution and the contribution of non-massive stars to the
interstellar radiation field which heats the dust.  The $\mathrm{K_S}$
band image of M\ 33 was taken from the $2MASS$ Large Galaxy Survey
catalog in the NASA/IPAC Infrared Science Archive.  The image was
background subtracted using a planar fit to regions well away from the
galaxy.  The $\mathrm{K_S}$ band flux can be contaminated by a variety
of sources other than non-massive stars.  They are the nebular and
molecular emission lines \citep{hunt03}, and hot dust \citep{devost99}.
Observations of star clusters have found that star clusters can have
excess H-K colors due to these contributors by at most 0.5 magnitudes
\citep{buckalew05}.  We thus apply this conservative estimate of 0.5
magnitudes (40\%) on the errorbar for the stellar component of the
$\mathrm{K_S}$ flux.  

The ionizing flux from massive (OB) stars cannot be measured directly
because of severe Ly$\alpha$ extinction, but we may estimate the
measure of the number of OB stars by a combination of H$\alpha$ and
24$\mu \mathrm{m}$ warm dust continuum, as done by \citet{calzetti07} to
derive extinction-corrected star formation rates.  The H$\alpha$ image
is from \citet{hoopes00}, and the 24$\mu \mathrm{m}$ from the
\textit{Spitzer} archive.  The scale factor when adding H$\alpha$ and
24$\mu \mathrm{m}$ is 0.031, given by \citet{calzetti07} in order to
minimize the dispersion of the combined star formation rate to that
derived from Pa$\alpha$ flux.  The dispersion is 20\%, which we take to be the
uncertainty in the measure of ionizing flux.  Uncertainties from aperture photometry
were negligible compared to 20\%. Details of the data reduction and preparation
for the 24$\mu \mathrm{m}$ image is found in \citet{onodera10}.

The HII regions to perform aperture photometry on were searched from
the literature with spectroscopic Oxygen abundances \citep{vilchez88,
crockett06, magrini07, rosolowsky08}.  A box with $54^{\prime \prime}$
on the side and centered on the HII region coordinate, was used to
locate the flux weighted center of each of the images within that box.
This centroid pixel was used as the center of the photometric
apertures.  This is because dust peaks are not always coincident
with the H$\alpha$ or $\mathrm{K_S}$ peaks, typically by
several to $\sim 10^{\prime \prime}$ , corresponding to under 40 parsecs.  All regions were then
inspected by eye to ensure that the measured emission was associated
with the HII region.  

  Flux from a total of 57 regions were measured in each of the 1.1\ mm,
  24 $\mu \mathrm{m}$, 160$\mu \mathrm{m}$, H$\alpha$ and $\mathrm{K_S}$ band images
 each using a circular aperture with a radius of $36^{\prime \prime}$.
The cold dust temperature was derived from the 1.1\ mm, 160$\mu \mathrm{m}$
using integrated fluxes from the photometry.  No local background was subtracted using an annuli in the 
photometry process, because most of the sources measured here are both crowded and extended.
An aperture correction factor of 1.745
, taken from the MIPS Instrument Handbook Version 2.0, was applied to the 160 $\mathrm{\mu m}$ photometry, for 30 K dust with no sky annulus subtraction.

Figure \ref{T-K-SFR} shows the relation between cold dust temperature,
$\mathrm{K_S}$ band, and the measure of ionizing flux.  A clear
correlation is seen between the cold dust temperature T and
$\mathrm{K_S}$ flux.  The correlation coefficient $r^2$ is 0.71.
Conversely, the correlation between T and ionizing flux is less
evident, with $r^2 = 0.26$.  Uncertainties in aperture photometry or temperature
are not sufficient to explain the difference in the dispersion. 
The fact that ionizing flux from massive
stars do not correlate with the dust temperature implies that these
stars are not responsible for heating the dust observed here, whereas stars
seen in the correlated $\mathrm{K_S}$ band are a strong candidate for dust heating.
Although flux from the Rayleigh-Jeans tail of the ionizing stars can potentially contribute also to the
$\mathrm{K_S}$ band, they cannot be the major dust heating source in
this case, because if so the correlation should be strong both in
$\mathrm{K_S}$ and the ionizing flux.  Similarly, the contribution of diffuse H$\alpha$ and 24$\mathrm{\mu m}$ emission to the
photometry, likely due to energy input from evolved stars \citep{verley07}, cannot explain a stronger correlation
between cold dust temperature T and $\mathrm{K_S}$ flux, compared to the combined H$\alpha$ and 24$\mathrm{\mu m}$.

This lends strong support to non-ionizing stars as the heating source of cold dust, even in the small scale star forming regions
discussed here.

\section{Summary and Conclusions}

This paper presented a large scale 1.1\ mm mapping observation of M\ 33, the
most proximate face-on spiral galaxy in the Local universe.  The
1.1\ mm continuum is detected out to $\sim 7$ kpcs, where the edge of
the star forming disk lies.
 By comparing the 1.1\ mm data with Spitzer MIPS 160$\mu \mathrm{m}$
imaging, we obtained the color temperature distribution of cold dust,
without significant contribution from warm dust as in most previous studies.
Furthermore, aperture photometry was performed on star forming
regions to compare the cold dust temperature with $\mathrm{K_S}$ band
flux, representing radiation from non-massive stars, and
H$\alpha$+$24\mu \mathrm{m}$ flux, representing ionizing radiation from
massive stars.  The following results are found.

\begin{itemize}
\item
The 1.1\ mm continuum is spatially correlated with star forming regions
     as traced by H$\alpha$, and comprise a clear spiral structure.

\item
The temperature of cold dust shows a smooth distribution over the entire
     disk.  The concentration of 1.1\ mm
     flux observed near the star forming regions are the result of more
     dust mass concentration in these regions, not a higher dust temperature.

\item
The temperature of cold dust shows a radially declining gradient, even
     though the detections are restricted to near the star forming regions.
     This is unlikely explainable instead by a gradient in the dust emissivity index $\beta$.

\item
A comparison in individual star forming regions with an aperture
     $36^{\prime \prime}$ in radius, or $\sim 300$ pc in diameter, shows
     that the cold dust temperature is correlated strongly with the
     local $\mathrm{K_S}$ band flux.  On the other hand, the temperature
     does not show a marked correlation with the ionizing flux, measured
     by combining H$\alpha$ and 24$\mu \mathrm{m}$ flux.  This shows
     that the heating source of cold dust near the star forming regions
     is not caused by massive ionizing stars, but the non-ionizing, evolved
     population of stars.\\[0.5cm]

\end{itemize}

We acknowledge R. Walterbos for providing us with the H$\alpha$ image.  We
are grateful to N. Ikeda, Y. Shimajiri, M. Hiramatsu, T. Minamidani,
T. Takekoshi and K. Fukue for the initial testing of the data reduction
pipeline, and S. Onodera for valuable discussions.  We thank N. Ukita
and the ASTE and AzTEC staff for the operation and maintenance of the
observing instruments, and the anonymous referee for valuable comments which
helped improve the paper.  The ASTE project is lead by Nobeyama Radio
Observatory, in collaboration with the University of Chile, the
University of Tokyo, Nagoya University, Osaka Prefecture University,
Ibaraki University and Hokkaido University.  S.K. was supported by the 
Research Fellowship from the Japan Society for the Promotion
of Science for Young Scientists.  This work is based in part
on archival data obtained with the NASA Spitzer Space Telescope.




\begin{figure}
   \includegraphics[width=10cm,keepaspectratio]{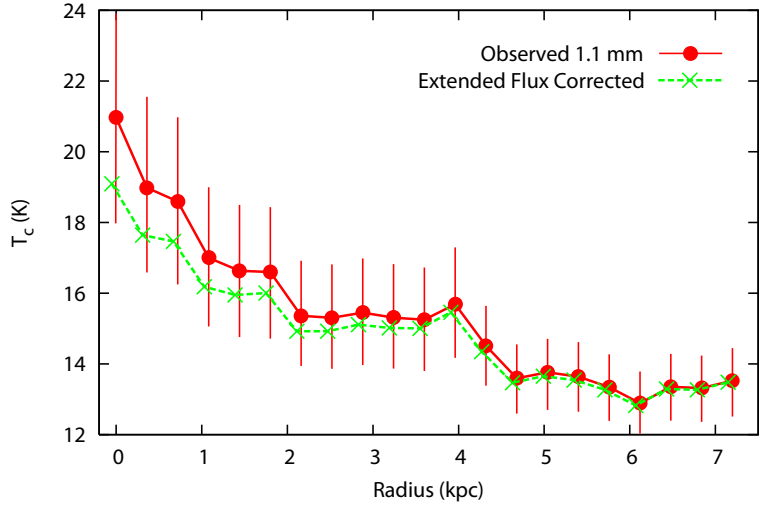}
  \caption{The cold dust temperature gradient.  Red circles are the temperature derived from observed
flux ratios.  Green crosses are temperatures after the correction for extended flux loss, explained in section 4.4.  Errorbars are determined from flux uncertainties in the observations (see section 4.3).
 \label{gradient}}
\end{figure}

\begin{figure}
   \includegraphics[width=10cm,keepaspectratio]{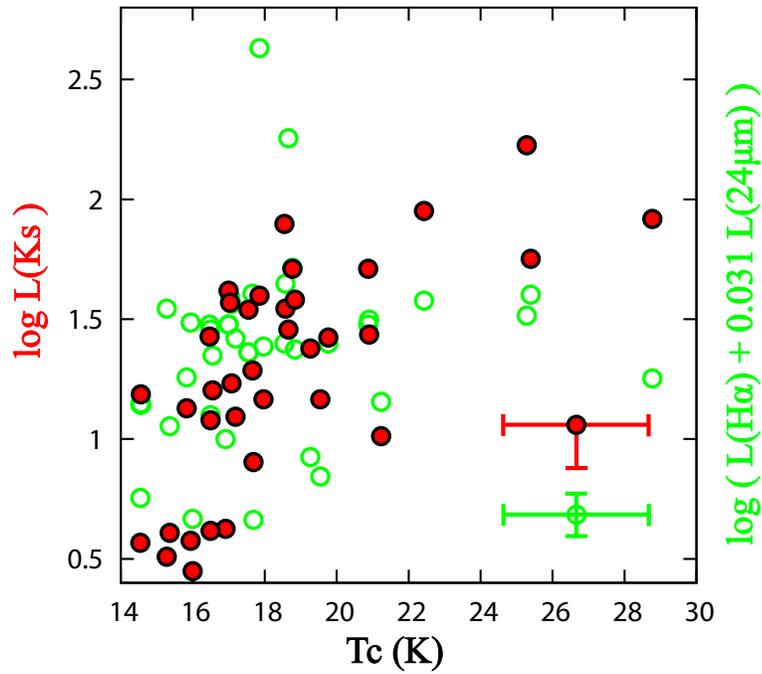}
  \caption{Relation between the cold dust temperature and $\mathrm{K_S}$
 flux (Red circles) and ionizing flux (Green circles) estimated from a
 combination of H$\alpha$ and 24$\mu \mathrm{m}$ flux.  The y-axis units
 are arbitrary shifted for comparison.  The errorbar for the
 $\mathrm{K_S}$ flux shows that the flux is given only as an upper limit,
 considering the contribution of hot dust or nebular/molecular lines.
 A uniform uncertainty of 2 K is used for the cold dust temperature
 errorbar, which is the characteristic value in figure \ref{gradient}.
 \label{T-K-SFR}}
\end{figure}

\begin{table}
\begin{center}
\caption{M\ 33 \label{m33param}}
\begin{tabular}{lcc}
\hline \hline
Parameter & Value & Reference$^{a}$ \\
\hline
 R.A. (J2000)   &  $1^\mathrm{h}33^\mathrm{m}50.9^\mathrm{s}$ & NED$^{b}$ \\
 Dec.           &  $30^{\circ}39^{\prime}36^{\prime \prime}$  & NED \\
 Inclination    &  $51^\circ$                                 & (1) \\
 Position Angle &  $22^\circ$                                 & (2) \\
 Morphology     &  SA(s)cd                                    & (3) \\
 Optical Size   &  $70.8^{\prime} \times 41.7^{\prime}$       & NED \\
 Distance       & $840\ $ kpc                                 & (4) \\
$\mathrm{M_B}$$^{c}$ &  $-19.38\ $mag.             & (5) \\ 
\hline
\end{tabular}

Notes $^{a}$ : (1) \cite{deul87}, (2) \cite{regan94}, 
(3) The Third Reference Catalogue of Bright Galaxies (RC3;\cite{rc3}),
 (4) \cite{freedman91}, (5) The HyperLEDA database, \cite{paturel03}.
$^{b}$ : NASA Extragalactic Database.
$^{c}$ : Absolute B band magnitude.
\end{center}
\end{table}

\clearpage

    \begin{table}
    \caption{Test sources $^{a}$} \label{embed}
\begin{center}
    \begin{tabular}{ccccc} \hline \hline
     R.A. & Dec. & FWHM & Peak Intensity & Total Flux \\
     J2000 & J2000 & $^{\prime \prime}$ & $\mathrm{mJy \ beam^{-1}}$ &     Jy \\  \hline
     $01^\mathrm{h}33^\mathrm{m}31.9^\mathrm{s}$ & $+30^\circ 56^{\prime}16.8^{\prime \prime}$ & $60.0$   & 10.0 & 0.058 \\
                                                 &                                             & $59.7$ & 9.6 & 0.055 \\
     $01^\mathrm{h}33^\mathrm{m}46.3^\mathrm{s}$ & $+30^\circ 59^{\prime}48.1^{\prime \prime}$ & $60.0$ & 15 & 0.087 \\
                                                 &                                             & $49.5$ & 15.9 & 0.062 \\         
     $01^\mathrm{h}34^\mathrm{m}10.9^\mathrm{s}$ & $+30^\circ 58^{\prime}54.9^{\prime \prime}$ & $120$ & 15.0 & 0.35 \\
                                                 &                                             & $102$ & 12.7 & 0.21 \\
     $01^\mathrm{h}34^\mathrm{m}55.9^\mathrm{s}$ & $+30^\circ 58^{\prime}42.8^{\prime \prime}$ & $120$ & 25.0 & 0.58 \\
                                                 &                                             & $110$ & 22.1 & 0.43\\
     $01^\mathrm{h}34^\mathrm{m}55.7^\mathrm{s}$ & $+30^\circ 50^{\prime}11.1^{\prime \prime}$ & $180$ & 45 & 2.3 \\ 
                                                 &                                             & $157$ & 36.2 & 1.4 \\ \hline
     \end{tabular}

Notes $^{a}$ : The lower row for each source corresponds to parameters obtained
 after reduction using the FRUIT algorithm.  The reduced data were
 fitted with two dimensional gaussian, centered at the position were it
 was embedded.
\end{center} 
\end{table}

    \begin{table}
\begin{center}
\caption{Flux Density at Various Wavelengths. \label{sedflux}}
    \begin{tabular}{ccc} \hline \hline
    Wavelength        & Flux & Reference$^{a}$  \\
     $\ \mu \mathrm{m}$ & $\log \mathrm{Jy}$ & \\ \hline
     24               & 1.70  & (1) \\
     60               & 2.75  & (2) \\
     70               & 2.82  & (1) \\
     100              & 3.10  & (2) \\
     160              & 3.31  & (1) \\
     170              & 3.34  & (2) \\
     1100             & 1.00  & This work  \\ \hline
     \end{tabular}

Notes $^{a}$ : (1): \citet{hinz04}. (2):\citet{hippelein03}.
\end{center}
\end{table}

\end{document}